\documentstyle[12pt,aasms4]{article}
\begin{document}

\title{Spherical Harmonic Expansion of Gamma Ray Burst Distributions:
Probing Large Scale Structure?}
\author{Tsvi Piran \altaffilmark{1}\authoremail{Tsvi@shemesh.fiz.huji.ac.il} \\
	and\\
Anupam Singh \ \altaffilmark{2}\authoremail{singh@tpau.physics.ucsb.edu}}
\altaffiltext{1}{ The Racah Institute for Physics, 
The Hebrew University, Jerusalem, Israel}
\altaffiltext{2}{Department of Physics, 
University of California, Santa Barbara, CA 93106, U.S.A. }

\def\prref{\par\noindent\hangindent=0.3cm\hangafter=1}
\def \APJ {{\it Ap. J.}}
\def \MN  {{\it M.N.R.A.S.}}
\def\kms{$ km \, sec^{-1}$}
\def\Mpc{\, h^{-1} \, {\rm Mpc}}
\def \etal {{\it et al.} }
\def \etalp {{\it et al.,} }

\def \eg { {\it e.g.} }
\bigskip

\hfill UCSBTH--96--16, July , 1996, Santa Barbara. 

\begin{abstract}
We have laid down the formalism and techniques necessary for computing
the multipole components in a spherical harmonic expansion for
bursting sources given any specific power spectrum of density
perturbations. Using this formalism we have explicitly computed and
tabulated the expected first few multipole components for the GRB
distribution using a Cold Dark Matter power spectrum.  Unfortunately,
our analysis leads us to expect that an anisotropy signal for this
model of structure formation will be below the shot noise level for
the foreseeable future.  Earlier studies by others had claimed that it
may be possible within a decade or so to probe structure formation
using the angular distribution of GRBs.  We find that while it may be
possible to probe the dipole due to our motion with respect to the
cosmic rest-frame, it does not seem feasible to probe the multipole
components due to intrinsic fluctuations if the actual power spectrum
is fixed to within an order of magnitude by the Cold Dark Matter model
of structure formation.

\end{abstract}

\section{Introduction}

The isotropy and paucity of weak bursts suggest that Gamma-ray bursts
(GRBs) are most likely cosmological.  Analysis of the number vs. peak
flux distribution (Cohen \& Piran, 1995; Fenimore \etalp 1993) 
show that the bursts originate from $z\approx 1$. This
suggests that GRBs could be applied as a tool for studying cosmology
and the large scale structure of the Universe.  If GRBs are
cosmological, they constitute a homogeneous population with a narrow
luminosity function (the peak luminosity of GRBs varies by less than
factor of 10 (Cohen \& Piran, 1995; Horack \& Emslie, 1994; Horack,
Emslie \& Meegan, 1994; Ulmer, Wijers \& Fenimore, 1995; Ulmer \&
Wijers, 1995) that is located at relatively high redshifts (Piran,
1992; Mao, \& Paczy\'nski, 1992; Wickramasinghe \etalp 1993;
Petrosian, 1993; Cohen \& Piran, 1995).  The universe and our Galaxy
are transparent to MeV range $\gamma$-rays (see e.g. Zdziarski \&
Svensson, 1989). Hence GRBs constitute a unique uniform population
which does not suffer from any angular distortion due to absorption by
the galaxy or by any other object. Additionally, GRBs are sampled
almost uniformly over $4 \pi$ steradians. The information on matter
fluctuations in the Universe is divided to local information obtained
from red-shift surveys that span distances up to $z\approx 0.05$ or
less and to information from the CMBR fluctuations that originate at
$z\approx 1000$.  The recent finding of Kollat \& Piran (1996)
that GRBs are correlated with Abell clusters suggest that GRBs do
follow the matter distribution and hence they could provide a unique
information about the distribution of matter at $z\approx 1$ and on
scales that, at present, cannot be explored in any other way.

Recently,   Lahav, Piran \& Treyer (1996) developed a scheme for
estimating the expected spherical harmonics that describe the
fluctuations in the intensity of the x-ray background. We modify this
scheme for estimating the spherical harmonics in the number of
bursting sources and apply the modified scheme to the GRB
distribution.  There are two sources for such fluctuations:
fluctuations in the number of sources that arises from fluctuations in
the matter density or from other effects such as the Sachs-Wolfe
effect or the Compton-Getting effect due to the motion of the earth
and random fluctuations that arises from Poisson noise.  We estimate
the expected fluctuations in the GRBs' angular distribution that
arise from both sources and we compare these theoretical estimates with
recent and future observations. The outline of this paper is the
following: In section 2. we develop the general formalism for
estimating the spherical harmonics in a general population of bursting sources
whose density fluctuates proportionally to the fluctuation in the
matter density. Using this formalism we estimate in section 3. the
number of GRBs needed in order that the real angular deviations of the
GRB population will be larger than the Poisson noise. In section
4. we estimate the fluctuations due to the Sachs-Wolfe effect and the
Compton-Getting effect.  In section 5. we compare our results with
previous works and with observational estimates.

\section{The Formalism}

We consider a general population of bursting sources that originate 
at cosmological distances and follow the matter distribution. The formalism 
is general but since we apply it to gamma-ray bursts (GRBs) we call the 
sources for simplicity GRBs in the rest of the discussion. 
We expand the number counts of GRBs in a given direction
${\bf \hat r}$ over the sky, $\sigma({\bf \hat r})$ 
in spherical
harmonic expansion:
\begin{equation}
\sigma({\bf \hat r}) = \sum_{lm}  a_{lm} Y_{lm}({\bf \hat r})
\end{equation}
where 
\begin{equation}
a_{lm} = \sum_{sources}  \; Y_{lm}^*({\bf \hat r_i})
\label{e2}
\end{equation}

To compare the observed distribution with a  theoretical one
We first estimate the number of observed bursts in a given direction
${\cal N}({\bf \hat r})$.
To do so we must made several assumption on the nature of the distribution of
GRBs. We first assume  that there is a linear biasing
between the  density  of GRB sources and the mass fluctuation 
\begin{equation}
\delta_x(r_c,\hat r) = b_{\gamma} \; \delta(r_c,\hat r),
\label{e3}
\end{equation}
where $b_{\gamma}$ is the bias factor between the density
perturbations and the GRB sources, $\delta ({\bf r})
\equiv  { {\delta \rho} / \rho } ({\bf r})$ is the 
density perturbation and $r_c$ is the comoving distance.  We also
assume that the rate of bursts per comoving volume per unit proper
time evolves with redshift as $(1+z)^p$. At present $p$ is unknown
(see e.g. Cohen \& Piran, 1995).  Finally, we must consider the
detection procedure of the specific detector. BATSE, for example, is
sensitive to $C$ the number of photons that arrive on a collecting area
$A$, within a given period of time, $\Delta T$, in a given energy
interval centered around $\bar E$:
\begin{equation}
C= { L \over 4 \pi r_c^2 (1+z)^{1+\alpha}} \; {A \Delta T \over 
\bar E} ,
\label{countrate}
\end{equation}
where $\alpha$ is the slope of the spectrum.

In general there exists a sensitivity function,
$\Phi(C)$, that describes the detectability of a burst with 
a count rate $C$.  Generally,
$\Phi(C) = 1 $ for $C\gg C_{min}$ and $\Phi(C) = 0 $
for $C\ll C_{min}$, where $C_{min}$ is the critical sensitivity of the
detector.
We will usually 
assume for simplicity that the detector behaves like a 
Heaviside function  and 
$\Phi(C) = 1 $ for $C\ge C_{min}$ and $\Phi(C) = 0 $
for $C < C_{min}$.
In this case  there will be a maximal comoving distance
$r_{c,max}$ from which a burst is detected for which 
$C(r_{c,max})=C_{min}$.

Using this assumptions we find that:
\begin{equation}
{\cal N}({\bf \hat r}) 
= \int_0^{\infty} dL \Phi(L)
\int r_c^2 dr_c  \; n (1+z)^{(p-1)} [1 + b_{\gamma} 
\delta(r_c, {\bf \hat r})] \Phi[C(L,r_c)]  ,
\label{N}
\end{equation}
where $r_c$ is the comoving distance, $\Phi(L)$ is the GRB luminosity 
function,
$n$ is an overall normalization such that for an observation period
$T_{obs}$, $n (1+z)^p/T_{obs}$ equals the effective density 
(per unit comoving volume per unit observer time) of bursts.
Eq. \ref{N} differs from the corresponding equation
for the x-ray background (Lahav, Piran \& Treyer, 1996) in two
ways. First the x-ray  background equation, in which we estimate the fluctuations
in the flux, is weighted by the flux from a given source. There is no such
weight here since we are interested in numbers of sources. Second, there is
additional factor of $(1+z)^{-1}$ in this equation since we are interested in
bursting sources, while for the x-ray background we were considering steady
state sources. The predicted harmonic coefficients are:
$$
a_{lm} = \int d {\bf \hat r} {\cal N}({\bf \hat r}) Y_{lm}^*({\bf \hat r})=
$$
\begin{equation} 
\int d{\bf \hat r} \int_0^{\infty} dL \Phi(L)
\int r_c^2 dr_c \; n (1+z)^{(p-1)} [1 + b_{\gamma} 
\delta(r_c, {\bf \hat r})] \Phi \left[r_{c,max}(L)-r_c\right]  \; 
Y_{lm}^*({\bf \hat r}),
\label{ee4}
\end{equation}

If $N$ is the total number of observed bursts,
we can use  the monopole ($l=0$), with $Y_{00} = (4 \pi)^{-1/2}$,  
to fix the normalization factor $n$ in the following way:
\begin{equation}
N =  ({4 \pi}) ^{1/2}  a_{00}  .
\end{equation}
For the special case of standard candles, $\Phi(L) = \delta (L-L_0)$,
with no source evolution $(p=0)$ we have
\begin{equation}
N= { 4 \pi r_{c,max}^3 \over 3} \left[ 1+ {3 H_0^2 r_{c,max}^2 \over {20 c^2}} 
- {3 H_0 r_{c,max} \over {4 c}} \right] n ,
\label{Nn}
\end{equation}
where $H_0$ is the Hubble constant and $r_{cmax}$ is the maximal
comoving distance from which a GRB with luminosity $L_0$ can be
detected.  Thus, for an Einstein-de Sitter universe we have:
\begin{equation}
n = {3 N \over{32 \pi}} {\left(H_0 \over c \right)}^3 { 1 \over 
{\left[ 1 - {1 \over {\sqrt{1+z_{max}}}} \right]^3}} { 1 \over {\left[ 0.1 + 
{0.6  \over {(1+z_{max})}} + {0.3 \over {\sqrt{1+z_{max}}}} \right]}} ,
\label{e7}
\end{equation}
where $z_{max}(L)=z(r_{c,max}(L))$ is the maximal redshift from which
a GRB with luminosity L can be observed.

The {\it fluctuations} in the background are expressed by higher
harmonics $l > 0 $:
\begin{equation}
a_{lm} = n \int dL \Phi(L) \int d {\bf \hat r}
\int r_c^2 d r_c \; \Phi[C(L,r_c)]  b_{\gamma} \delta ( r_c, {\bf \hat r})\; 
Y_{lm}^*({\bf \hat r}) {(1+z)}^{p-1}.
\label{e8}
\end{equation}
Expressing the fluctuations in the matter density  in terms of
the Fourier components $\delta_k$ (see Eq. \ref{A4})
and using  the orthogonality of Spherical Harmonics 
$\int d \omega Y_{lm} ({\bf \hat r}) Y_{l'm'}^* ({\bf \hat r}) = 
\delta_{ll'}^{mm'}$ we find:
\begin{equation}
a_{lm} =  n (i^l)^* { 1 \over { 2 \pi^2} }
\int d^3 k b_{\gamma} \delta_{\bf k} (z) Y_{lm} ({\bf \hat k} ) 
\int dL  \Phi(L) \int  r_c^2 dr_c \; \Phi[C(L,r_c)] j_l(kr_c) {(1+z)}^{p-1}.
\label{e9}
\end{equation}

We are interested in sources at $z\approx 1$ and at large wavelengths.
The relevant fluctuations that contribute to this integral  
are still in the linear regime.
This means that:
\begin{equation}
\delta_{\bf k} (z) =  \delta_{{\bf k}0}/(1+z) 
\label{e10}
\end{equation}
where $\delta_{{\bf k}0}$ is the present linear amplitude of
fluctuations {\it normalized to fit the CMB observations}.
Eq. \ref{e10} is gauge dependent (in some gauges perturbation that are
larger that the horizon do not grow). However, it can be checked
numerically that the contribution of very large scale perturbations is
small and hence this gauge choice is not important.

Taking the mean-square values and using Parseval's theorem and
\begin{equation} 
\langle \delta_{\bf k} (z) \; \delta_{\bf k'}^* (z) \rangle \; =  
(2 \pi)^3  P(k,z) \delta^{(3)}({\bf k} - {\bf k'} ) 
\label{e12}
\end{equation}
we obtain  
\begin{equation}
\langle |a_{lm}|^2 \rangle = { 32 \over \pi} \;  b_{\gamma}^2 ( { c \over H_0}
)^6 n^2  \int dk k^2  P(k) |\Psi_l(k)|^2 .
\label{e13}
\end{equation}  
The function $\Psi_l(k)$ is defined (for an Einstein -de Sitter
universe) as:
\begin{equation}
\Psi_l(k)  \equiv  \int dL \Phi(L) 
\int_0^{z_{max}}  dz  \;  (1+z)^{p - 7/2}\;
\left[ 1 - {1 \over {\sqrt{1+z}}} \right]^2  j_l(k r_c)  \Phi[C(L,z)]  .
\label{e14}
\end{equation}

\section{Estimates for GRBs}

We consider standard candles GRBs with no density evolution i.e. $p =
0$.  For simplicity we use in this section $\Phi(C)=1$ for $C \ge
C_{min}$ and $\Phi(C)=0$ otherwise.  The power-spectrum for CDM can be
written down in the form (Padmanabhan, 1993):
\begin{equation}
P(k) = { A k^n \over{ (1+ B k + C k^{3/2} + D k^2)^2 }} .
\label{e15}
\end{equation}
For a scale-invariant or Harrison-Zel'dovich spectrum, $n = 1$.
Further, by fitting to available data we get the best-fit values of
the parameters as follows: $ A = (24 Mpc)^4 $, $ B = 1.7 Mpc$, $C = 9
Mpc^{3/2}$ and $D = 1 Mpc^2$.  Let us scale out the dimensions of $k$
in the following way,
\begin{equation}
\tilde k \equiv  k (100 Mpc)
\label{e16}
\end{equation}
By doing this rescaling the expression for $\langle |a_{lm}|^2
\rangle$ can be written completely in terms of dimensionless numbers
in the following form:
\begin{equation}
\langle |a_{lm}|^2 \rangle = {   (3.0 \times 10^{-5})   N^2 
\over{ \left[ 1 - {1 \over {\sqrt{1+z_{max}}}} \right]^6} 
{\left[ 0.1 + {0.6 \over {(1+z_{max})}} + {0.3 \over
{\sqrt{1+z_{max}}}} \right]^2}}
\;
b_{\gamma}^2
\int d\tilde k { \tilde k^3 |\Psi_l(k)|^2 \over {D(\tilde k)}}
\label{e17}
\end{equation}
where 
\begin{equation}
D(\tilde k) = (1 + 0.017 \tilde k + 0.009 \tilde k^{3/2} + 0.0001
\tilde k^2)^2 .
\label{e18}
\end{equation}
Thus, the quantity which needs to be estimated numerically is:
\begin{equation}
\tilde I (l,z_{max}) \equiv \int d\tilde k { \tilde k^3 |\Psi_l(k)|^2 
\over {D(\tilde k)}}
\label {e19}
\end{equation}

Let us now obtain an order of magnitude estimate of the above effect
and see whether this is going to be detectable in the near future
(Tegmark \etalp 1995; 1996).  Let us re-write the earlier equation
\ref{e17} in a form convenient for tabulation of values for various
values of $z_{max}$
\begin{equation}
\langle |a_{lm}|^2 \rangle = {  p(z_{max})  N^2} \;
b_{\gamma}^2  \tilde I(l,z_{max})
\label{26}
\end{equation}
Here $p(z_{max}) $ is the numerical pre-factor which depends on $z_{max}$,
which we shall tabulate below.\\

TABLE 1 \\

\begin{tabular}{|r||c|c|c|c|} \hline
$z_{max}$ & 0.2 & 0.5 & 1.0 & 2.0 \\ \hline $p(z_{max})$ & $9.0 \times
10^1$ & 1.4 & $1.3 \times 10^{-1}$ & $2.4 \times 10^{-2}$ \\ \hline
\end{tabular} 
\vspace{0.1 in}\\

One can estimate the magnitude of $\tilde I (l,z_{max})$ for various
values of $l$.\\

TABLE 2 \\

\begin{tabular}{|r||c|c|c|c|} \hline
$z_{max}$ & 0.2 & 0.5 & 1.0 & 2.0 \\ \hline $\tilde I (1,z_{max})$ &
$3.09 \times 10^{-9}$ &$ 8.20 \times 10^{-9} $ & $1.064 \times
10^{-8}$ & $1.036 \times 10^{-8}$ \\ \hline $\tilde I (2,z_{max})$ &
$3.90 \times 10^{-9}$ &$ 1.061 \times 10^{-8}$ & $1.439 \times
10^{-8}$ & $1.487 \times 10^{-8}$ \\ \hline $\tilde I (3,z_{max})$ &
$4.72 \times 10^{-9}$ &$ 1.310 \times 10^{-8}$ & $1.826 \times
10^{-8}$ & $1.949 \times 10^{-8}$ \\ \hline $\tilde I (4,z_{max})$ &
$5.51 \times 10^{-9}$ &$ 1.559 \times 10^{-8}$ & $2.214 \times
10^{-8}$ & $2.411 \times 10^{-8}$ \\ \hline $\tilde I (5,z_{max})$ &
$6.29 \times 10^{-9}$ &$ 1.804 \times 10^{-8}$ & $2.598 \times
10^{-8}$ & $2.868 \times 10^{-8}$ \\ \hline
\end{tabular} 
\vspace{0.1 in}\\

We, of course have to compare the magnitude of the signal given in the
previous equation to the shot noise. The magnitude of the shot noise
in our case is given by :
\begin{equation}
\langle |a_{lm}|^2 \rangle_{sn}  = 
{1 \over 4 \pi} \int dV_c {n \over {(1+z)}}  = 
{N \over {4 \pi} } 
\label{e24}
\end{equation}   
where $N$ is the number of bursts observed.  Let us now estimate what
is the number of bursts one needs to get to observe a signal above
noise. This means we want,
\begin{equation}
 {p(z_{max}) N^2  } \;
b_{\gamma}^2  \tilde I(l,z_{max}) >  {N \over {4 \pi} }  
\label{e25}
\end{equation}

We can obtain an estimate for the number of bursts required to get a
signal above noise for any desired assumption about the location of
the most distant sources of GRBs.  In particular let us obtain the
required number of bursts if we assume the most distant sources of
GRBs is at $z_{max} = 0.2$.  We start at Eq. \ref{26} and use the
value of $p(z_{max}) $ given in Table 1 to get the relation,
\begin{equation}
\langle |a_{lm}|^2 \rangle = { N^2 (9.0 \times 10^{1}) } \; b_{\gamma}^2  \tilde
I(l,0.2) \label{e27}
\end{equation}
One can numerically compute the magnitude of $\tilde I(l,0.2)$ for
various values of $l$. We did this for l = 1 to l = 5.  The magnitude
does not change significantly and is given by, $\tilde I \simeq 4
\times 10^{-9} $.  Thus the relevant inequality now becomes,
\begin{equation}
 {  N^2 (9.0 \times 10^{1})  } \;
b_{\gamma}^2  \tilde I(l,0.2) >  {N \over {4 \pi} }  
\label{28}
\end{equation}
Thus if the most distant GRBs are at $z_{max} = 0.2$ then the number
of bursts required to get a signal over noise must be given by $ N > 2
\times 10^5$.  In general, if the most distant GRBs are at $z_{max}$
then the number of bursts required to get a signal over noise must be
given by $ N > N_c(l,z_{max}) b_{\gamma}^{-2}$, where

\begin{equation}
N_c(l,z_{max}) = \left[4 \pi {p(z_{max})   } \;
\tilde I(l,z_{max})\right]^{-1}  
\end{equation}

Below we tabulate the Number of bursts, $N_c (l,z_{max})$ required to
observe the multipole component $l$ for a given maximum redshift of
sources $z_{max}$:\\

TABLE 3 \\

\begin{tabular}{|r||c|c|c|c|} \hline
$z_{max}$ & 0.2 & 0.5 & 1.0 & 2.0  \\ \hline
$N_c (1,z_{max})$ & $2.9 \times 10^{5}$   &$ 6.9 \times 10^{6} $ & $5.8 \times 10^{7}$ & $3.2 \times 10^{8}$  \\ \hline
$N_c (2,z_{max})$ & $2.3 \times 10^{5}$   &$ 5.4 \times 10^{6}$ & $4.3 \times 10^{7}$       & $2.2 \times 10^{8}$  \\ \hline
$N_c (3,z_{max})$ & $1.9 \times 10^{5}$   &$ 4.3 \times 10^{6}$ & $3.4 \times 10^{7}$       & $1.7 \times 10^{8}$  \\ \hline
$N_c (4,z_{max})$ & $1.6 \times 10^{5}$   &$ 3.6 \times 10^{6}$ & $2.8 \times 10^{7}$       & $1.4 \times 10^{8}$  \\ \hline
$N_c (5,z_{max})$ & $1.4 \times 10^{5}$   &$ 3.2 \times 10^{6}$ & $2.4 \times 10^{7}$       & $1.2 \times 10^{8}$  \\ \hline
\end{tabular} 
\vspace{0.1 in}\\

These results are perhaps discouraging because it means that the
expected signal is not detectable in the foreseeable future. However,
given the fact that there are ongoing efforts to measure the multipole
components of the GRB distributions (Tegmark \etalp 1995;1996), this
is still a significant result.  It means that if a signal is detected
it is not due to the effect discussed here by us and an alternative
explanation must then be sought for the signal.

\section{Sachs-Wolfe and Compton-Getting}

We turn now to fluctuations in the redshift that influence the
observed count rate and induce fluctuations in the number density
given by Eq. (\ref{N}).  So far we have introduced $n = n_0 + \delta
n$ where $n_0$ is the background density and $\delta n$ are the
fluctuations and we were interested in those fluctuations (we assumed
that $\delta n = b_{\gamma} n_o \delta \rho$ where $b_{\gamma}$ is a
biasing factor and $\delta \rho$ the fluctuations in the matter
density.  Now we consider fluctuations in the count rate $C$ and in
the redshift factor $(1+z)$ that arise due to Doppler motions of the
sources and the observer $(V_0-v_{||})$ (where $V_0$ is our motion and
$v_{||}$ is the motion of the source), or due to fluctuations in the
gravitational potential $\delta \phi$.  For simplicity we discuss in
this section a standard candle population.  The generalization to a
luminosity function is trivial.

The variation in the factor of $(1+z)$ causes fluctuations in the time
dilation between the sources and the observer and hence fluctuations
in the observed rate of events from different regions of
space-time. These lead to the following fluctuations in the angular
count rate:
\begin{equation} 
\delta {\cal N}_1 = \int r_c^2 dr_c {n (1+z)^{p-1}}  
[- (V_0-v_{||})/c - \delta \phi/3 c^2] .
\label{dN}
\end{equation}
The first term $V_0$ is a constant vector that gives rise, just like
in the CMBR, to the Dipole and a second order quadrupole term:
\begin{equation}
{ \Delta N_{d1} \over {\bar N} } 
= { V_{obs} \over c } \cos (\theta) + \bigg({ V_{obs} \over c }\bigg)^2
[1- \cos^2(\theta)/2] ~.
\label{dipole}
\end{equation}
Note that for sources that are at $z\approx 1$ the intrinsic
fluctuations discussed earlier (Eq. (21) and Tables 1 and 2) give a
dipole at the $10^{-4}$ level.  The Doppler dipole discussed above in
Eq. (28) is at a few times $10^{-3}$ level.  Thus the intrinsic dipole
is comparable to but about an order of magnitude below the Doppler
dipole. However, by examining the same equations and tables as above
we see that the intrinsic quadrupole is clearly much greater than the
Doppler quadrupole.  In general the higher Doppler multipole
components are suppressed by increasing powers of $\left(V_{obs} / c
\right)$ whereas the higher multipole moments of the intrinsic
fluctuations are of the same order as the lower ones (see Table 2).

The second and third terms, $ v_{||}$ and $\delta \phi$, reflect the
fluctuations of the velocity field and the gravitational field.  Like
in the discussion of the density fluctuations we turn to Fourier
space. Using:
\begin{equation}
\delta \phi_{\bf k} = {3 \over 2} { H^2 a^2 \over k^2 }  \delta_{\bf k}(z)
={3 \over 2} {H_0^2}  (1+z) { \delta_{\bf k}(z)  \over k^2}
\label{dphi1}
\end{equation}
for $\delta \phi$ and 
\begin{equation}
v_{||} =  { H a} {{\delta_{\bf k}}(z) \over k}
=  H_0 \sqrt{1+z} {{\delta_{\bf k}}(z) \over k}
\label{dvv}
\end{equation}
for the line-of-sight velocity $v_{||}$, we obtain:
\begin{equation}
\delta  \phi/c^2 = { 3 H_0^2 \over {2 c^2}} (1+z)  {1 \over {2 \pi^2 } }
\sum_{lm} (i^l)^* Y_{lm}^* ({\bf \hat r}) 
\int d^3k \; {\delta_{\bf k} \over
k^2} Y_{lm} ({\bf \hat k}) \; j_l(kr) 
\label{dphi} 
\end{equation}
and (Fisher, Scharf \& Lahav, 1994)
\begin{equation}
v_{||}({\bf r})/c  = {{ H a} \over
{2\pi^2 c}} \sum_{lm} (i^l)^* Y_{lm}^*(\hat{\bf r}) \int d^3k\, 
{{\delta_{\bf k}}(z) \over k}
Y_{lm}(\hat{\bf k}) j_l^{'}(kr) \ .
\label{vii} 
\end{equation}

Substitution of Eqs. \ref{dphi} and \ref{vii} into \ref{dN} yields:
\begin{equation}
\delta {\cal N}_1 ({\bf \hat r})
= {1 \over {2 \pi^2 } }  \int r_c^2 dr_c  n   (1+z)^{p-1} 
 \sum_{lm} (i^l)^* Y_{lm}^* ({\bf \hat r}) 
\label{dN1} 
\end{equation}
$$ 
\bigg [
{H_0^2 \over 2}   \int d^3k \; {\delta_{\bf k} \over
{k^2 c^2}} Y_{lm} ({\bf \hat k}) \; j_l(kr) 
+ { H_0  \over \sqrt{1+z} }   \int d^3k\, 
{{\delta_{\bf k}}(z) \over {k c}} Y_{lm}(\hat{\bf k}) j_l^{'}(kr) 
\bigg]~.
$$ To calculate the corresponding $ a_{lm} $ we expand $\delta {\cal
N} ({\bf \hat r})$, expressed in Eq. \ref{dN1} :
\begin{equation}
a_{lm,1} = { (i^l)^*  \over {2 \pi^2 } } n 
\int d^3k \; Y_{lm} ({\bf \hat k}) \delta_{\bf {k0}}  
\bigg[{H_0^2 \over {2k^2 c^2}} \Psi_{l,sw} + {H_0  \over {k c }}  \Psi_{l,v}
\bigg]~, \label{almf}
\end{equation}
where
\begin{equation}
\Psi_{l,sw}(k)  \equiv 4 ({c \over H_0})^3  
\int_0^{z_{max}}  dz  \;  (1+z)^{p - 5/2}\;
\left[ 1 - {1 \over {\sqrt{1+z}}} \right]^2  j_l(k r_c)~,
\label{psil}
\end{equation}
and
\begin{equation}
\Psi_{l,v}(k)  \equiv 4 ({c \over H_0})^3  
\int_0^{z_{max}}  dz  \;  (1+z)^{p - 4}\;
\left[ 1 - {1 \over {\sqrt{1+z}}} \right]^2  j'_l(k r_c)~.
\label{psiv}
\end{equation}

%
%

The fluctuations in the gravitational field and in the velocity field
also influence the detectability of bursts that are marginally
detectable and this leads to another effect.  The BATSE detector is
sensitive to $C$ the number of photons in a given energy interval that
arrive to the detector in a given period of time.  The count rate $C$
varies due to variation in z which arise from the Doppler effect and
the gravitational redshift.  Using Eq. \ref{countrate} we find:
\begin{equation}
{\partial  C \over \partial z}
=  -{ (1+ \alpha) C \over (1+z)}~.
\label{dCz}
\end{equation}

Substitution of Eq. \ref{dCz} to Eq. \ref{N} and linearizing we obtain:
\begin{equation}
\delta{\cal N}_{2} = \int_0^{\infty}{  n (1+z)^{p-1}} 
{\partial \Phi\over \partial C} (1+ \alpha) C
{[-(V_0-v_{||})/c - \delta \phi/3c^2]}  r_c^2 dr_c ~.
\label{deltaNphi}
\end{equation}

For simplicity we assume that our detector has a sharp cutoff at
$C_{min}$.  Generalization to a smooth cutoff it trivial. In this case
$\Phi$ is a Heaviside function, $ {\partial \Phi\over \partial C} =
\delta(C - C_{min}) $ and the integral becomes a surface term!

We can express the volume element as a Jacobian times $dC$.  Thus, we
arrive at the following trivial integral for $ \delta {\cal N}_{2}$:
$$
\delta {\cal N}_{2} = \left({c \over {H_0} } \right)\int_0^{\infty}{  
n\over (1+z)}  \delta(C - C_{min})
{\left[(V_0-v_{||})/c + \delta \phi/3c^2\right]} r_c^2 {1 \over
{\sqrt{1+z}}} d C = $$
\begin{equation}
\left({c \over {H_0} } \right){  n\over (1+z_{max})^{3/2}} 
{\left[(V_0-v_{||}(z_{max}))/c  + \delta \phi(z_{max})
/3c^2\right]} r_c^2(z_{max})~.
\label{dNphi}
\end{equation}
The quantities: $v_{||}(z_{max})$, $\delta \phi(z_{max})$ and
$r_c^2(z_{max})$ are all calculated on the surface $z_{max}$ which
corresponds to $C_{min}$ with our standard candle luminosity.

The contribution from $V_0$ is again a dipole and a second order 
quadrupole term:
\begin{equation}
{ \Delta N_{d2} \over {\bar N} } 
={(V_{obs}/c) \cos (\theta) + ( V_{obs}/c )^2
[1- \cos^2(\theta)/2] \over 
{\left[ \sqrt{1+z_{max}}- 1 \right]} {\left[ 0.1 (1+z_{max})+
 {0.3  {\sqrt{1+z_{max}}}}
+0.6 \right]}}~. 
\label{dipole2}
\end{equation}
We discuss the magnitude of these terms as well as those given
in Eq. \ref{dipole} in the next section.

The fluctuation terms can be dealt in the same manner that we have
used earlier.  First, we substitute Eqs. \ref{dphi} and \ref{vii}
into Eq. \ref{deltaNphi}.  Then, after multiplying by $Y_{lm}(\hat r)$
and integrating over $d\Omega$ we get:
\begin{equation}
a_{lm,2} = { (i^l)^*  \over {2 \pi^2 } } n
 \int d^3k \; Y_{lm} ({\bf \hat k}) \delta_{\bf {k0}}
\bigg[  { H_0^2 \over {2k^2 c^2}}  \Psi_{l,sw,S} +
{ H_0  \over {k c }}  \Psi_{l,v,S} \bigg]~,
\label{alms}
\end{equation}
where:
\begin{equation}
\Psi_{l,sw,S}(k)  \equiv 4 ({c \over H_0})^3
 (1+z_{max})^{p - 3/2}\;
\bigg[1- {1 \over \sqrt{1+z_{max}}} \bigg]^2
j_l(k r_c)~,
\label{psilS}
\end{equation}
and
\begin{equation}
\Psi_{l,v,S}(k)  \equiv 4 ({c \over H_0})^3
 (1+z_{max})^{p - 3}\;
\bigg[1- {1 \over \sqrt{1+z_{max}}} \bigg]^2j^{'}_l(k r_c)~.
\label{psivS}
\end{equation}

%
%

We can combine now all fluctuating terms (Eqs. \ref{e9},
\ref{almf} and \ref{alms}) 
to a single equation 
for $a_{lm}$  which resembles Eq. \ref{e9} with 
$\Psi_l$ replaced by 
$\Psi_l + \Psi_{l,sw}+\Psi_{l,sw,S} +\Psi_{l,v} + \Psi_{l,v,S}$:
$$
a_{lm} = { (i^l)^*  \over {2 \pi^2 } } n
 \int d^3k \; Y_{lm} ({\bf \hat k}) \delta_{\bf {k0}}
\bigg[b_\gamma \Psi_l +  
$$
\begin{equation}
{  H_0^2 \over {2k^2 c^2}} (\Psi_{l,sw}+
\Psi_{l,sw,S}) +
{ H_0  \over {k c}}  (\Psi_{l,v}+ \Psi_{l,v,S} ) \bigg]
\label{almfull} ~.
\end{equation}
Now, we continue and
calculate $\langle a_{lm} a^*_{lm}\rangle$   taking 
the mean-square values and using Parseval's theorem and Eq. \ref{e12}:
$$
\langle | a_{lm} |^2 \rangle = {32 \over \pi} \left(c \over H_0 \right)^6 n^2 
\int dk k^2 P(k) \left| \bigg[ b_\gamma \Psi_l + \right.
$$
\begin{equation}
{ H_0^2 \over {2k^2 c^2}} 
(\Psi_{l,sw} + \Psi_{l,sw,S})  +
\left. { H_0  \over {k c}} (\Psi_{l,v} + \Psi_{l,v,S} ) \bigg] \right|^2~.
\label{almav}
\end{equation}

%
%

>From the above expressions we can calculate the most general r.m.s.
amplitude $\langle a_{lm} a^*_{lm}\rangle$. It is worthwhile,
however, to estimate the order of magnitude of the relevant terms.
To do so we recall that if
$\lambda$ is the length scale being probed then (Padmanabhan, 1993):
\begin{equation}
\delta \phi/c^2 \simeq \left({\delta \rho \over \rho} \right)_\lambda
\left( H~ \lambda \over c \right)^2 ,
\label{dv}
\end{equation}
and
\begin{equation}
(V_0-v_{||})/c  \simeq \left({\delta \rho \over \rho} \right)_\lambda
\left( H~ \lambda \over c \right) \; .
\label{xx}
\end{equation}
These estimates suggest that the Sachs-Wolfe terms in Eq. \ref{almfull}
are smaller than the intrinsic fluctuation term by a factor
of $(H\lambda /c)^2$. The velocity terms are smaller than the 
intrinsic fluctuation by a factor of $H \lambda /c$. 

Comparison of the surface terms in the Sachs-Wolfe and the 
velocity terms to the volume integral show that these terms
are comparable for small $z_{max}$ and become smaller as
$z_{max}$ becomes larger. This is intuitively explained by
the fact that at lower $z_{max}$ there are relatively more
faint bursts than at higher $z_{max}$.  

>From our expressions it is clear that there is a systematic
expansion in terms of the parameter $\left( {H_0 /{k c}} \right)$.
These results, show that the Sachs-Wolfe and the 
velocity terms can be ignored
if one is interested in the lowest order effect and an order of
magnitude calculation as we have been interested here. 
The difference between GRBs and CMBR is that
the first arise from sources at relatively low
red-shift, for which the intrinsic fluctuations have grown and
are much more important than the fluctuations induced by a varying
gravitational field or from the random peculiar velocity of 
the sources.

\section{Discussion}

We have laid down the formalism and techniques for evaluating the
expected multipole components of the GRB count distribution.  The
largest term, which is a dipole of order $10^{-2}$, almost two orders
of magnitude larger then all other terms arises due to the
Compton-Getting effect resulting from our motion relative to the GRB
distribution.  This term is not new.  It was first calculated by Maoz
(1994) who estimated to be of order $\approx 10^{-2}$. Later Scharf
\etal (1995) repeated the calculations and have considered also the
fluence weighted dipole (which we don't consider here).

Smaller multipole components arise from intrinsic fluctuations of the
source distribution that are related to the power spectrum of density
perturbations.  We have explicitly computed the first few multipole
components for a Cold Dark Matter power spectrum. These are in fact
the very multipole components that are least likely to be affected by
the positional inaccuracy in locating GRBs. We feel that the
expression of the fluctuations in the GRB distribution in terms of a
multipole moments is particularly useful in view of the general usage
of these moments to express fluctuations in the CMBR (Bond \&
Efstathiou, 1987) and other background fields (Peebles, 1973; Lahav,
1994; Fisher, 1993; Heavens \& Taylor, 1994; Lahav, Piran \& Treyer,
1996).  There have now been serious and exhaustive efforts (Tegmark 
\etalp 1995; 1996) to examine the deviation from isotropy of GRBs
using an expansion in terms of spherical harmonics. Further, this
observational program is likely to continue into the future and so a
calculation of the expected signal in terms of various multipole
components is a useful calculation to do.

Our work would certainly be incomplete if we failed to mention the
earlier work of Lamb \& Quashnock (1993) who were the first to address
the issue of whether it is possible to probe Large Scale Structure
using the distribution of Gamma Ray Bursts, and did not compare our
results to their analysis and conclusions.  LQ use the angular
two-point correlation function to examine the expected deviations from
isotropy, whereas we have used an expansion in terms of spherical
harmonics. While undoubtedly both formalisms have their usefulness we
have already mentioned that we do feel that an expansion in terms of
spherical harmonics is better suited to address the issue of
deviations from isotropy.  However, having said all that in favor of
doing a spherical harmonic expansion of the GRB distribution it still
remains a true statement that such conclusions as the number of bursts
required to see an anisotropy signal over the noise should be basis
independent.

We now turn to comparing our results about the number of bursts
required to see a signal above noise to those obtained by LQ. The
central observational quantity in the formalism that LQ use is the
mean of the angular two-point correlation function, ${\bar
{w}}(\theta)$. The statistical uncertainty in ${\bar{w}}(\theta)$ is
denoted by $\delta{\bar{w}}(\theta)$. In order to detect a signal it
is therefore necessary to require $\delta{\bar{w}}(\theta) /
{\bar{w}}(\theta) < 1$.  The above inequality can be transformed into
an inequality which determines the minimum number of bursts required
to detect a signal due to the actual anisotropy of the distribution as
opposed to the expected statistical fluctuations. This bound is
written down by LQ as
\begin{equation}
N_b \geq {{2 \pi^2 {\sqrt{8}} D^3}\over{C I(k_o) P(k_o)}} ~,
\end{equation}
where $N_b$ is the number of bursts, $D$ is the distance to the edge
of the source distribution, $k_o$ is the largest wavenumber that can
be probed, $P(k)$ is the power spectrum of density perturbations,
$C$ is a constant introduced by LQ and $I(k_o)$ is the following integral
introduced by LQ,
\begin{equation}
I(k_o) =  \int^{k_o} {{dk}\over{k}} (kD)^2 W(k) ~.
\end{equation}
Here $W(k)$ is a window function introduced by LQ. While LQ are
somewhat ambiguous about the exact numerical value of the window
function used to arrive at their results, they do argue that in the
regime $D^{-1} < k < (D \theta)^{-1}$ the window function can be computed 
analytically tending for large values of $kD$ to ${5.64 \over{kD}}$.
Further, they argue that most of the contribution to the integral,
$I(k_o)$ comes from wavenumbers $k \sim k_o$. Using these inputs we
obtain an order of magnitude estimate for $I(k_o)$ to be 
\begin{equation}
I(k_o) \simeq  5.64 k_o D  ~.
\end{equation}

We also need to input the power spectrum to extract the bound on the number
of bursts for a detectable signal. To make a fair comparison let us do 
this along the lines suggested by Figure 1 of LQ. Thus we take,
$k_o = 0.1 h Mpc^{-1}$ and $P(k_o) \simeq 5 \times 10^2 h^3 Mpc^3$.
LQ also argue that $C \simeq 1.5$ in order for their convolution to
match those from Limber equation for small values of $\theta$.
Further they choose $D = 1 h^{-1} Gpc$ for illustrative purposes.
Note that this corresponds to $z_{max} \simeq 0.3$. Putting in all these
values we get the bound on the number of bursts, $N_b$ for a detectable
signal to be,
\begin{equation}
N_b \geq 2 \times 10^5 ~. 
\end{equation}
This result agrees in order of magnitude with our results for
comparable $z_{max}$.
Thus we see that within the context of our analysis both the results of the
LQ formalism and the spherical harmonic expansion agree. However, there
is one claim made by LQ with which we disagree.
Namely, they claim that if an anisotropy signal is not detected with
$ \sim 3000$ bursts then it would cast a doubt on the cosmological origin
of GRBs. Such a statement we think is unwarranted. Certainly if one stays
within the context of the Cold Dark Matter models whose spectrum we have used
in our analysis and the maximum redshift from which GRBs originate
$z_{max} \geq 0.2$ there would be no reason to expect an anisotropy signal
above that of shot noise with only $ \sim 3000$ bursts.

Let us now summarize our results in the light of the observational
situation.  We first note that BATSE has accumulated $\sim 1000$
bursts in 3 years of operation. If it continues operating for a period
of 10 years it will accumulate $\sim 3000$ bursts. In light of our
results displayed in Table 3 this means that if GRBs originate at
redshifts $z_{max} \geq 0.2$, and the Cold Dark Matter power spectrum
is right to within an order of magnitude, then we are unlikely to see
an anisotropy signal in GRB distributions above the shot noise within
the next 10 years or so.  Given that there are now serious and
exhaustive efforts to measure and analyze the spherical harmonic
components of the GRB distribution (Tegmark \etalp 1995; 1996), we
think an accurate estimate of the signal expected from Large Scale
Structure was necessary.

We have laid down in this paper the formalism and techniques necessary
for computing the various multipole components in a spherical harmonic
expansion for bursting sources given any specific power spectrum of density
perturbations. We have further explicitly computed and tabulated
the first few multipole components for GRBs distribution using
a Cold Dark Matter power spectrum.
Unfortunately, our analysis leads us to expect that an anisotropy signal
for this model of structure formation 
will be below the shot noise level for the foreseeable future.

\section{Acknowledgments}

We would like to thank the  participants and organizers of the
ITP workshop on Gamma Ray Bursts for informative talks and a 
stimulating program. We would like to thank  Ofer Lahav,
David Langlois, Reem Sari,  Omer Blaes, Richard Holman and 
Mark Srednicki for 
stimulating discussions on related topics. This work was supported 
in part by the U.S. NSF Grant PHY--91--16964, and by the Israeli NSF.

\newpage
\appendix{\bf Appendix: Fourier Preliminaries}
\vskip 0.2 in
Fourier decomposition of the density perturbations 

\begin{equation} 
\delta ({\bf r}) \equiv 
{ {\delta \rho} \over \rho } ({\bf r}) =   
{1 \over ({2 \pi}) ^3 } \int d^3k \; \delta_{\bf k} \;
 e^{- i {\bf k}\cdot {\bf r}} 
\label{A1}
\end{equation}
and the inverse:
\begin{equation}
\delta_{\bf k} = \int d^3 r \; \delta({\bf r}) \; e^{i {\bf k}\cdot 
{\bf r}} ~.
\label{A2}
\end{equation}
Rayleigh expansion:
\begin{equation}
e^{i {\bf k}\cdot {\bf r}} = 4 \pi \sum_{lm} i^l j_l(kr) 
Y_{lm} ({\bf \hat r}) 
Y_{lm}^* ({\bf \hat k}) ~.
\label{A3} 
\end{equation}
Combining Eqs. (\ref{A1}) and (\ref{A3}) : 
\begin{equation}
\delta ({\bf r}) = 
{1 \over {2 \pi^2 } } \sum_{lm} (i^l)^* 
Y_{lm}^* ({\bf \hat r}) 
\int d^3k \; \delta_{\bf k} Y_{lm} ({\bf \hat k}) \; j_l(kr) ~.
\label{A4} 
\end{equation}

\end{document}